\documentclass[11pt]{article}
\pdfoutput=1
\usepackage{amsmath,epsf,amssymb,latexsym,enumerate,cite,shadow,array,color}
\usepackage{slashed}
\usepackage{graphicx,wasysym}

\DeclareFontFamily{U}{rsf}{} \DeclareFontShape{U}{rsf}{m}{n}{
  <5> <6> rsfs5 <7> <8> <9> rsfs7 <10-> rsfs10}{}
\DeclareMathAlphabet\Scr{U}{rsf}{m}{n} \makeatletter
\def\be{\begin{equation}}
\def\ee{\end{equation}}
\def\ba{\begin{array}}
\def\ea{\end{array}}

\newcommand{\beq}{\begin{equation}}
\newcommand{\eeq}[1]{\label{#1}\end{equation}}
\newcommand{\bea}{\begin{eqnarray}}
\newcommand{\eea}[1]{\label{#1}\end{eqnarray}}

\def\Re{\mathop{\rm Re}\nolimits}
\def\Im{\mathop{\rm Im}\nolimits}

\usepackage{ifpdf}
\ifx\pdfoutput\undefined
   \pdffalse
\else
   \pdfoutput=1
   \pdftrue
  \usepackage[pdftex]{hyperref}
\begin{document}

\begin{titlepage}
\hfill CERN-PH-TH/2014-133\\

\hskip 1.5cm

\begin{center}

{\huge \bf{

Minimal $R+R^2$ Supergravity Models of Inflation Coupled to Matter}
 }

\vskip 0.8cm  

{\bf \large Sergio Ferrara$^{1,2,3}$ and Massimo Porrati$^4$}  

\vskip 0.5cm

\noindent $^{1}$ Physics Department, Theory Unit, CERN, CH 1211, Geneva 23, Switzerland\\
$^{2}$ INFN - Laboratori Nazionali di Frascati, Via Enrico Fermi 40, I-00044 Frascati, Italy\\
$^{3}$ Department of Physics and Astronomy, University of California Los Angeles, CA 90095-1547 USA\\
$^{4}$ {CCPP, Department of Physics, NYU
4 Washington Pl. New York NY 10003, USA}

\end{center}

\vskip 1 cm

\begin{abstract}
The supersymmetric extension of ``Starobinsky" $R+\alpha R^2$ models of inflation is particularly simple in the ``new minimal" formalism of  supergravity, where the inflaton has no scalar superpartners. This paper is devoted to matter couplings in such supergravity models. We show how in the new minimal formalism matter coupling presents certain features absent in other formalisms. In particular, for the large class of matter couplings considered 
in this paper,  matter must possess an R-symmetry, which is gauged by the vector field which becomes dynamical in the ``new minimal'' completion of the $R+\alpha R^2$ theory. Thus, in the  dual formulation of the theory, where the gauge vector is part of a massive vector multiplet, the inflaton is the superpartner of the massive vector of a nonlinearly realized R-symmetry. The F-term potential of this theory is of no-scale type, while the inflaton potential is given by the D-term of the gauged R-symmetry. The absolute minimum of the potential is always exactly supersymmetric, so in this class of models if realistic vacua exist, they must be always metastable. We also briefly comment on possible generalizations of the examples discussed here and we exhibit some features of higher-curvature supergravity  coupled to matter in the ``old minimal" formalism.

 \end{abstract}

\vspace{24pt}
\end{titlepage}



\section{Introduction}

In this paper we analyze the coupling to matter of a new form of supergravity inflationary theory proposed in~\cite{Ferrara:2013rsa} in a special yet interesting case. In the absence of matter,
such special case is the supergravity embedding of the bosonic $R+\alpha R^2$~\cite{Starobinsky:1980te,Kofman:1985aw,Whitt:1984pd} gravity discussed in~\cite{cfps}. 

In the presence of matter, two situations must be distinguished, depending on whether a superpotential exists. 
These situations are discussed in sections 2 and 3. The main point is that in the ``new minimal" auxiliary field formulation, the supergravity auxiliary vector $A_\mu$ is a gauge field, whose gauge symmetry requires matter to be coupled in a specific way. For zero superpotential, any K\"abler potential is allowed, since matter fields have zero conformal weight. However, in the presence of a superpotential, the ``new minimal" formalism requires the presence of a gauged R-symmetry, that is a symmetry under which the superpotential 
has weight 3.

In the presence of the $R^2$ term, the auxiliary field $A_\mu$ becomes dynamical and massive, and the original $U(1)_{SC}$ of the superconformal symmetry is spontaneously broken. Indeed, the new minimal $R+\alpha R^2$ theory is equivalent after a superfield duality to a standard ``old minimal" supergravity with a massive vector multiplet. Note that such massive vector multiplet, whose action is described by a real function $J(C)$~\cite{vp}, has its own Fayet-Iliopoulos (FI) term --a term linear in $C$ in $J(C)$-- which is responsible for the almost de Sitter plateau during inflation. Adding to such system R-symmetric matter results in a particular standard supergravity where the vector multiplets containing $A_\mu$ gauges the R-symmetry. The resulting scalar Lagrangian is described in section 4. Matter coupling is described by a real function of the scalar fields $z$, $\Phi(z,\bar{z})$, while the canonically normalized
inflaton is identified with the following combination of fields defined in the next sections: 
$\exp(\sqrt{2/3}\phi)=T+\bar{T}- \Phi(z,\bar{z})/3$. Remarkably, the potential has a no-scale structure~\cite{noscale}. 

In section 5, the scalar field dynamics for a rather generic theory dual to $R+\alpha R^2$ coupled to matter is described. We observe that the matter potential is almost identical to that of a globally supersymmetric theory, as is to be expected because of the no-scale structure. This fact suggest that a realistic supersymmetry breaking vacuum, if it exists, can at most be metastable, as in the scenario explored in the context of global supersymmetry in~\cite{iss}.

Possible extensions of our construction are discussed in section 6. One natural generalization is to change the self-interactions of the massive vector to a generic K\"ahlerian $\sigma$-model, which is no longer an $SU(1,1)/U(1)$ coset with fixed curvature, as in the case of the $R+\alpha R^2$ theory. Another possibility is to add $R^n$ corrections similarly to what was done in ref~\cite{fklp2}. In this case, the supersymmetric completion of these new terms will depend on the scalar combination defined in this paper: $T+\bar{T}- \Phi(z,\bar{z})/3$. Section 6 also comments on higher scalar curvature supergravity coupled to matter in the ``old minimal" formalism.

\section{Coupling to Matter in the Absence of a Superpotential}

The key feature of ref.~\cite{cfps} is the use of the ``new minimal" formulation of supergravity. This formulation uses a linear multiplet, $L$, of conformal weight $c=2$ and chiral weight $w=0$, as conformal compensator.

Our first observation is that, in  the absence of superpotential, a quite general form of the Lagrangian of the $R+\alpha R^2$  theory coupled to matter is
\beq
{\cal L}=-[LV_R +L\Phi(z,\bar{z})/3] _D + {1\over 2g^2}[W_\alpha (V_R) W^\alpha(V_R)]_F +c.c., \qquad V_R\equiv \log (L/S \bar{S}).
\eeq{m1}
The term $-LV_R$ gives rise to the standard Einstein action while the $R^2$ term comes from the kinetic term of the real  superfield $V_R$; the parameters $\alpha$ and $g$ are related by $\alpha=1/2g^2$. The chiral supefield $S$ is the compensator of  ``old minimal'' supergravity. The function $\Phi$ of the $c=w=0$ chiral superfields $z,\bar{z}$ is a real superfield of zero chiral and conformal weights, but it is otherwise arbitrary. The only constraint comes from the fact that $\Phi$ is the K\"ahler potential of the $z$ scalars~\footnote{This is different from the old minimal case, where the function of chiral superfields that multiplies the
compensators in the D-term is the {\em exponential} $\exp(-K/3)$, with $K$ the K\"ahler potential.}. Notice that Lagrangian~(\ref{m1}) is independent of $S$, thanks to the gauge invariance of the F-term and the property $[L(\Omega+\bar{\Omega})]_D=0$, which holds for any liner multiplet $L$ and chiral multiplet $\Omega$. 

A convenient way to transform eq.~(\ref{m1}) into a standard supergravity form is to use a real-multiplet Lagrange multiplier $B$~\cite{Ferrara:2013rsa} (see also ref.~\cite{Farakos:2013cqa}) in order to rewrite Lagrangian~(\ref{m1}) in terms of an unconstrained vector multiplet $U$ and the redundant chiral compensator $S$ as
\beq
{\cal L}=-[S\bar{S}e^U (U +\Phi(z,\bar{z})/3)] _D + [B(S\bar{S}e^U-L)]_D+ {1\over 2g^2}[W_\alpha (U) W^\alpha(U)]_F +c.c.,
\eeq{m2}
Lagrangian~(\ref{m2}) reduces to~(\ref{m1}) upon using the $B$ equation of motion. 
Using the $L$ equation of motion one finds instead $B=T+\bar{T}$, with $T$ chiral. 
A redefinition of the compensator $S$, namely $S\equiv S_0 e^{-T}$  casts eq.~(\ref{m2}) into the standard {\em old minimal} form
\beq
{\cal L}=-[S_0 \bar{S}_0 e^{U-T-\bar{T}} (U -T - \bar{T} +\Phi(z,\bar{z})/3)]_D + 
{1\over 2g^2}[W_\alpha (U) W^\alpha(U)]_F +c.c. \, .
\eeq{m4}
Notice that $T$ appears as the St\"uckelberg supermultiplet of the nonlinearly realized symmetry gauged by $U$:
\beq
T\rightarrow T+\Omega, \qquad \Omega=\mbox{chiral superfield} .
\eeq{m3}
The matter fields have zero charge under this symmetry; their K\"ahler potential is 
\beq
K=-3\log[T+\bar{T} -\Phi(z,\bar{z})/3] +3T +3\bar{T}.
\eeq{m5}
Naturally, the term $3T+3\bar{T}$ can be canceled by a K\"ahler transformation in the absence of a superpotential, but the same transformation adds a constant to the D-term of the $U$ gauge field. 
\section{Adding a Superpotential} 

Since the fields $z,\bar{z}$ have zero chiral and conformal weight, the superpotential appears in eq.~(\ref{m1}) as the F-term  $[W(z)S^3]_F+c.c.$. An explicit dependence on the chiral compensator
$S$ is forbidden in the new minimal scheme, where the only compensator is the linear multiplet $L$.
A solution to this problem --not the most general, but general enough for the purpose of this paper--  is to consider superpotentials of weight 3 under and Abelian symmetry, which leaves the kinetic function $\Phi$ invariant. So from now on we will select functions $W(z^I)$, $\Phi(z^I,\bar{z}^I)$ such that there exist charges $q_i$ with the property
\beq
W(e^{iq_I \theta } z^I)=e^{3i\theta}W(z^I), \qquad \Phi(e^{iq_I\theta}z^I, e^{-iq_I\theta}\bar{z}^I)=
\Phi(z^I,\bar{z}^I), \qquad \mbox{for } \theta \in \mathbb{R}.
\eeq{m6}
This {\em R-symmetry} can be promoted to a gauge symmetry of the supergravity Lagrangian by introducing a vector superfield $V$ and defining the gauged R-symmetry by
\beq
V\rightarrow V+ \Omega + \bar{\Omega}, \qquad z_I \rightarrow e^{q_I\Omega}  z_I, \qquad 
S \rightarrow e^{-\Omega}S, \qquad \Omega=\mbox{ chiral superfield}.
\eeq{m7}
The problem is how to fit this symmetry into our $R+\alpha R^2$ theory. The correct way turns out to be that $V$ has to be identified with $V_R$, the ``composite'' real field that contains the scalar curvature $R$. 
The invariant Lagrangian can now be written as follows 
\beq
{\cal L}=-[S\bar{S}e^U(U +\Phi(e^{-q_IU}z_I,\bar{z}_I)/3)] _D + [B(S\bar{S}e^U-L)]_D
+{1\over 2g^2}[W_\alpha (U) W^\alpha(U)]_F + [S^3W(z)]_F + c.c. \, .
\eeq{m8}
By solving the $B$ equations of motion we find as before that $S\bar{S}e^U$ is a linear multiplet. 

Lagrangian~(\ref{m8})  does not depend on $S$, in spite of the presence of a superpotential, because this field can be changed to $Se^{-A}$, with $A$ {\em any} chiral superfield, by using the redefinition 
$ U\rightarrow U-A - \bar{A}, \; z^I \rightarrow e^{-q_IA}  z^I$, $B\rightarrow B- A - \bar{A}$. 

As in section 2, by solving the $L$ equations of motion and setting $S=S_0e^{-T}$ we arrive at the standard {\em old minimal} supergravity Lagrangian
\beq
{\cal L}=-[S_0 \bar{S}_0 e^{U-T-\bar{T}}(U -T -\bar{T} +\Phi(e^{-q_IU}z^I,\bar{z}^I)/3)] _D +{1\over 2g^2}[W_\alpha (U) W^\alpha(U)]_F + [S_0^3e^{-3T}W(z^I)] + c.c. \, .
\eeq{m9}
The superpotential of the theory, $e^{-3T}W(z^I)$ has overall weight zero under the symmetry gauged by $U$. The possibility of redefining R-symmetry into a standard ``matter" symmetry thanks to a St\"uckelberg field, when R-symmetry is broken everywhere in field space, was noticed in~\cite{Catino:2011mu}. Moreover, since the symmetry is always broken, any potential anomaly can be canceled by a 4D Green-Schwarz mechanism~\cite{Green:1984sg}.
\section{Scalar Field Lagrangian}

The potential and kinetic terms of the scalars are easily derived from eq.~(\ref{m9}) using  general formulas given in~\cite{cfgvp} and~\cite{Ferrara:1983dh}. The latter reference is relevant because it includes the coupling of a gauged 
R-symmetry, that is a symmetry under which the superpotential has a nonzero weight. To compute the scalar action we define the function $G$, which combines the 
K\"ahler potential and the superpotential $e^{-3T}W$ as
\beq
G= K+ \log (e^{-3T}W )+ \log (e^{-3\bar{T}}\bar{W}) = -3\log[T+\bar{T} -\Phi(z,\bar{z})/3] +\log W +\log \bar{W} .
\eeq{m10}
Supergravity practitioners will immediately recognize a no-scale structure~\cite{noscale} in this equation, but we nevertheless point out some steps in deriving the scalar potential for all other readers~\footnote{A no scale structure exists also in the K\"abler potential of the ``old minimal" formulation of $R+\alpha R^2$ 
supergravity~\cite{Cecotti:1987sa},
but not in its superpotential, see eq.~(\ref{m20}). This is due to the Jordan frame origin of the inflaton in all supersymmetric completions of
$R+\alpha R^2$ gravity. }.

We use the formulas for F-term potential and D-terms given in~\cite{cfgvp,Ferrara:1983dh}. 
The scalar potential is generically a sum of two terms: $V=V_F+ V_D$. The second contribution is absent when no gauge fields couple to matter. The first is 
$V_F=e^G (G_AG^{AB}G_B-3)$, where $G_A\equiv \partial G/\partial y^A$ etc. and $y^A=(T,z^I)$.
$G^{AB}$ is the inverse of $G_{AB}$. The following quantities --together with their complex conjugate--  are useful for deriving $V_F$:
\bea
&& G_T=-3/X , \qquad G_I=\Phi_I/X + W_I/W , \nonumber \\ &&
 G_{T\bar{T}}=3/X^2, \qquad  G_{I\bar{T}}=-\Phi_I/X^2, \qquad 
 G_{I\bar{J}}=\Phi_{I\bar{J}}/X +\Phi_I \Phi_{\bar{J}}/3X^2 .
 \eea{m11}
 Here we defined $X=T+\bar{T} -\Phi/3$. 
 It is also useful to exploit the definition of $G^{AB}$, which gives the following relations, together with their complex conjugates
 \beq
 3G^{T\bar{T}}-\Phi_I G^{I\bar{T}}=X^2, \qquad 3G^{T\bar{I}} -\Phi_J G^{J\bar{I}}=0, \qquad
  -\Phi_I G^{\bar{T} J} +(G_{I\bar{L}}X+ \Phi_I \Phi_{\bar{L}}/3) G^{\bar{L}J}=X^2 \delta_I^J.
 \eeq{m12}
 Finally, using the definition $\Phi^{I\bar{L}}\Phi_{L\bar{J}}=\delta^I_J$ we have
 \beq
 G^{I\bar{J}}=X \Phi^{I\bar{J}}.
 \eeq{m13}

By exploiting eqs.~(\ref{m10}-\ref{m13}) and after a short calculation, the F-term contribution to the potential
is found to be
\beq
V_F= {1\over X^2} W_I\Phi^{I\bar{J}} \bar{W}_{\bar{J}} .
\eeq{m14}

As expected because of the no-scale structure, this potential is identical with the one of rigid supersymmetry up to the
rescaling by the factor $1/X^2$. 
The D-term potential comes from the gauge superfields $U$~\footnote{Of course many more 
D-terms exist in a realistic system, one for each generator of the gauge group of matter; these do not change the 
properties of the potential during slow roll but may be relevant after the end of inflation. It is anyway straightforward to add them to our potential.} and reads
\beq
D_U/g^2=2G_T + \sum_I q_Iz^I G_I + q_I\bar{z}^I G_{\bar{I}}= -6/X+
\sum_I q_Iz^I \Phi_I /X+ q_I\bar{z}^I \Phi_{\bar{I}}/X +6,
\eeq{m15}
where in the last equation we used the property that the matter superpotential is a homogeneous function of degree
3: $\sum_I q_Iz^IW_I=3W$~\cite{Ferrara:1983dh}.

The D-term contribution to the scalar potential is positive definite by construction. The term $D^2_U/2g^2$ is $O(g^2)$ and is the ``Starobinsky" inflaton potential discussed in~\cite{Farakos:2013cqa,Ferrara:2013rsa}.

Before commenting on the slow roll and post-inflation dynamics of the scalars, let us write the scalar kinetic term $K_{\mu\nu}g^{\mu\nu}$. Calling  as before $A_\mu$ the gauge vector in the supermultiplet $U$ and using standard supergravity manipulations~\cite{cfgvp} together with definitions~(\ref{m10},\ref{m11}) one finds
\bea
K_{\mu\nu}&=&{\Phi_{I\bar{J}}\over X}D_\mu z^I D_\nu z^{\bar{J}} + 
{3\over 4 X^2} \partial_\mu X \partial_\nu X + \nonumber \\ &&
 + {3\over X^2}[\partial_\mu \Im T +A_\mu- (\Im D_\mu z^I\Phi_I/3)]
[\partial_\nu \Im T +A_\nu - \Im (D_\nu z^I\Phi_I/3)],
\eea{m16}
with $D_\mu z^I\equiv \partial_\mu z^I +iq^IA_\mu z^I$.
The kinetic term shows that  the canonically normalized inflaton field $\phi$, with kinetic term $g^{\mu\nu}\partial_\mu \phi \partial_\nu \phi/2$, is $\phi = \sqrt{3/2}\log X$ and that as expected $\Phi_{I\bar{J}}$ must be positive definite.

\section{Scalar Field Dynamics}

A noticeable feature of $V_F$ in eq.~(\ref{m14})  is that up to the overall rescaling $1/X^2$, it is identical  to the
potential of a rigid supersymmetric theory without inflaton. This factor can be significant; in fact, the standard slow-roll relation between e-foldings and 
potentials (see e.g. \cite{Turner:1992cf}), $N=\int_{\phi_i}^{\phi_f} d\phi V/ (dV/d\phi) $ (in units where $8\pi G=1$), implies in our case $X\approx 4N/3 $ ($\approx 80$ for $N=60$). So, the F-term potential of matter is multiplied by a factor that during slow roll can be $O(10^{-4})$. This means that during slow roll the mass of all scalars is reduced by a factor $\sim 3/4N$, compared with their post-inflation value. So, many scalars can become so light during slow roll that they may change the slow roll dynamics. This is a model-dependent problem worth further study but beyond the scope of this letter. 

Here we will more modestly assume that the F-term potential is steep enough to rapidly settle all ``matter" fields $z^I$ to the minimum of the potential~(\ref{m14}): $W_I\sim 0$. One such point, which always exists when the theory is regular for small $z^I$ and the superpotential $W$ contains no linear terms in $z^I$, is $z^I=0$ for all $I$: this is the R-symmetric vacuum. Notice that in this vacuum the D-term contribution makes the potential become simply
\beq
V=2g^2(3/X -3)^2=18g^2 [\exp(-\sqrt{2/3}\phi )-1]^2;
\eeq{m17}
this is the "Starobinsky" potential studied in~\cite{Ferrara:2013rsa} (and many other places besides). 

When inflation ends at $X=O(1)$,  if we assume that the matter scalars are characterized by scales much below $H\sim g M_{Pl}$, then their dynamics is the standard post-inflationary one, where the inflaton is frozen at the $X=1$ minimum. If R-symmetry breaking minima occur at VEVs $|\langle z^I\rangle| \ll H$, then the possibility exists for matter to undergo a phase transition to a more realistic vacuum where both R-symmetry and supersymmetry are broken without exciting the inflaton away from its $X=1$ minimum. 
{\em Notice though that the supersymmetric vacuum will always be an absolute minimum of the potential}. So, to be 
phenomenologically viable, the class of models proposed here must realize supersymmetry breaking {\em \`a la} ISS~\cite{iss}.

\section{Generalizations and Comments}

In ref.~\cite{Ferrara:2013rsa}, the Starobinsky potential was just one among a large class of slow roll potentials that can be obtained by coupling a massive vector superfield $U$ to supergravity. In the absence of matter and
superpotential, the $U$ field coupling is defined by a function $J(U)$~\cite{vp,fayet}, related to the K\"ahler function of the 
St\"uckelberg field $T$ by $J(U-T-\bar{T})=-K(T +\bar{T} -U)/2$. In this paper, we studied the coupling to 
matter of the model defined by $J=(3/2)[\log (-U) +U]$. A straightforward generalization of our paper is 
to couple matter to a massive vector using an arbitrary function $I(U)$ of the vector field:
\beq
K=-3\log [ I(U-T-\bar{T} ) -\Phi(z,\bar{z})/3] +3T +3\bar{T} -3U.
\eeq{m17a}
The case studied in this paper is $I(U)=-U$. This is the case when the K\"abler manifold of the complex scalar $T$ defines, for
constant values of the matter fields $z^I$, the coset space $SU(1,1)/U(1)$, with fixed radius of curvature~\cite{Ferrara:2013rsa}. This and other nice features exhibited by our new minimal supergravity model arise because the interactions 
of the massive vector supermultiplet with matter are quite restricted. More general new minimal interactions, such as
 those considered in ref.~\cite{Ferrara:1983dh}, or those arising from the supersymmetric completion of the $R^n$ terms considered in~\cite{fklp2}, are worth studying and deserve a fuller investigation, that we hope to continue in future works. 
 
 We conclude with some brief comments on the coupling of higher scalar curvature supergravity to matter in the old minimal formalism. Such coupling assumes a particularly simple form when the Lagrangian of the theory can be written as
 \beq
 {\cal L}= -[(1-h({\cal R}/S,\bar{{\cal R}}/\bar{S})/3-\Phi(z,\bar{z})/3)S\bar{S}]_D + [W(z)S^3]_F +c.c. \, ,
 \eeq{m18}
 where ${\cal R}$ is the chiral superfield containing the scalar curvature in its $\theta^2$component: ${\cal R}= .. +\theta^2 R +...$. 
 The ``sequestered" form of matter interactions, whose Jordan function and superpotential simply add to the higher-curvature sector, are assumed here for simplicity, but they are perhaps justifiable in a higher-dimensional 
 setting as in~\cite{Randall:1998uk}.
 Thanks to the identity $[T{\cal R}S^2]_F + c.c. = [(T+\bar{T}) S\bar{S}]_D$, and following the steps in~\cite{Cecotti:1987sa}, Lagrangian~(\ref{m18}) can be shown to be dual to a standard supergravity theory with two additional chiral multiplets: 
 $T$ and $A$~\cite{fgvn}. 
 The dual Lagrangian is
\beq
{\cal L} = -[(T+\bar{T} -h(A,\bar{A})/3 -\Phi(z,\bar{z})/3)S\bar{S}]_D +[(W(z)-(T-1/2)A)S^3]_F +c.c. \, .
\eeq{m19}
Stability of the $A,\bar{A}$ scalars during inflation can be achieved by modifying the pure $R+\alpha R^2$ theory, defined by the function $h(A\bar{A})=\alpha A\bar{A}$, by adding to it higher scalar curvature terms such as $\zeta A^2 \bar{A}^2$ with $\zeta > 0.5$~\cite{Kallosh:2013lkr}. 

By standard manipulations and defining $(A,z^I)\rightarrow z^a$,  $h+\Phi =\Psi$ we find that the scalar potential has an almost no-scale structure (cfr.~\cite{Ellis:2013nxa}):
\bea
V&=&{1\over X^2} [ W_a\bar{W}_{\bar{b}} \Psi^{a\bar{b}} + W_T \bar{Y}_{\bar{T}} + \bar{W}_{\bar{T}}Y_T], \label{m20} \\
\bar{Y}_{\bar{T}}& =& - \bar{W} + {1\over 3}\bar{W}_{\bar{a}} \Psi^{\bar{a}b}\Psi_b  +{1\over 2} \bar{W}_{\bar{T}} \left(
{X\over 3} +{1\over 9} \Psi_a \Psi_{\bar{b}} \Psi^{a\bar{b}}\right), \label{m21} \\
 X&=&  T+\bar{T} -\Psi/3, \qquad \Psi^{a\bar{c}}\Psi_{\bar{c}b}=\delta^a_b.
\eea{m22}
If $W_T=0$ before and after inflation, the potential simplifies to
\beq
V={1\over X^2} \left( {1\over h_{A\bar{A}}}W_A \bar{W}_{\bar{A}} + W_I W_{\bar{J}}\Phi^{I\bar{J}} \right).
\eeq{m23}
The Starobinsky potential is the $\Re T$ part of the $|W_A|^2$ term; the same potential gives a mass to $\Im T$.
The potential~(\ref{m23}) looks identical to the new minimal potential, in which the D-term of the massive vector multiplet
is replaced by the F-term of the $A$ field. Setting $W_T=A=0$ is the same result~\cite{Antoniadis} that one would obtain by replacing the ``sgoldstino" superfield~\cite{Kallosh:2013lkr,Ferrara:2013wka} with a Volkov-Akulov fermion that realizes supersymmetry 
nonlinearly~\cite{Antoniadis,AlvarezGaume}. This condition can be realized in many ways, e.g. at the point $z^I=0$, if the matter superpotential $W(z^I)$ is at least quadratic in the $z^I$ fields.

 \subsection*{Acknowledgements} We  are grateful to R. Kallosh and A. Linde for stimulating collaborations on related previous investigations and to A. Kehagias for useful discussions.
 S.F.   is supported by ERC Advanced Investigator Grant n. 226455 {\em Supersymmetry, Quantum Gravity and Gauge Fields (Superfields)}.
M.P. is supported in part by NSF grant PHY-1316452. M.P. would like to thank CERN for its kind hospitality and the ERC Advanced Investigator Grant n. 226455 for support while at CERN.

\end{document}